\documentstyle[12pt]{article}

\def\beq{\begin{equation}} \def\eeq{\end{equation}}
\def\bea{\begin{eqnarray}} \def\eea{\end{eqnarray}}
\let\nn=\nonumber

\begin{document}
\baselineskip=17pt

{\large\rm DESY 93-085}\hfill{\large\tt ISSN 0418-9833}

{\large\rm June 1993}\hfill\vspace*{2cm}
\begin{center}
{\bf {\large \bf Searching for Supersymmetry in High-Energy\\
\bf Photon-Proton Scattering}} \\
\vspace{0.5truecm}
W. Buchm\"{u}ller, Z. Fodor\footnote{Humboldt Fellow,
on leave from Institute for Theoretical Physics,
E\"otv\"os University, Budapest, Hungary}\\
Deutsches Elektronen-Synchrotron DESY, Hamburg, Germany
\end{center}
\vspace*{2cm}

\begin{abstract}
We investigate the possibility to search for supersymmetry in the
scattering of protons and Compton back-scattered
laser light. We evaluate the cross sections for inelastic and
elastic production of wino pairs for different electron-proton c.m.s.
energies. For $\sqrt{s}=1\ TeV$ the cross section exceeds $1\ fb$
for $m_w<190\ GeV$.
\end{abstract}
\vfill\eject

It has been known for some time that in the minimal supersymmetric extension
of the standard model the running gauge couplings of strong, weak and
electromagnetic interactions tend to meet at a unification scale
$\Lambda_{GUT}$, whereas this is not the case in the ordinary
non-supersymmetric standard model \cite{amaldi}. Furthermore,
the unification scale $\Lambda_{GUT}$ is large enough in order
to increase the theoretical predictions for the proton lifetime
beyond the present experimental lower limit. Recent precision
experiments \cite{lep} have confirmed \cite{susygut} this observation.
The unification of gauge couplings in the supersymmetric standard
model appears even more remarkable if various experimental and theoretical
uncertainties are carefully taken into account \cite{langacker}.

In view of this success several calculations of the mass spectrum in the
supersymmetric standard model with radiative symmetry breaking
have recently been carried out, where constraints from the top-quark
mass and considerations on fine tuning of parameters were taken into account
\cite{mass}. A generic feature of the resulting mass spectrum is the ordering
\beq\label{ordering}
m_{\tilde \gamma},m_z,m_w<m_h,m_{\tilde l}<m_{\tilde q}<m_{\tilde g},
\eeq
for the mass of photino $(\tilde \gamma)$, zino $(z)$, winos$(w)$,
higgsinos $(h)$, scalar leptons $(\tilde l)$, scalar quarks $(\tilde q)$
and gluino $(\tilde g)$. The masses of the light gauginos $\lambda$,
$z$ and $w$ are typically of order $100\  GeV$ \cite{mass}.

What are the prospects to discover light gauginos in the $100\ GeV$ mass
range at present and future colliders? In the near future LEP II
offers the best chances. Here it will be possible to discover winos with
masses almost up to the electron beam energy, i.e. $100\ GeV$. For $e^+e^-$
machines with higher c.m.s. energies the discovery limit essentially scales
with the beam energy. Due to lower production cross sections and higher
background rates the discovery of electroweak gauginos
is difficult at hadron colliders.

Recently, in connection with linear $e^+e^-$ colliders, the interesting
possibility has been discussed to turn an electron beam into a hard photon
beam by means of Compton back-scattered laser light, and
intensive theoretical work has
already been carried out in order to clarify the
physics potential of $e\gamma$ and $\gamma\gamma$ machines \cite{telnov}.

Hence, one may also consider the discovery potential of high-energy
collisions between photons and protons. The protons are available at
present storage rings and the photons could be generated by use of laser
light, back scattered from electrons of a linear collider.
\footnote{The question, which luminosity one can reach in high-energy
photon-proton collisions, still has to be investigated.}
For instance,
with an electron beam of $250\ GeV$ and a $1\ TeV$ proton beam one would
reach a c.m.s. energy of $1\ TeV$. Since high energy protons yield a
considerable flux of high energy photons, the resulting two-photon
subprocess gives a sizeable cross section for wino pair production
(cf. fig.1).

Let us consider this process in more detail. The low energy
effective lagrangian for photons, W-bosons, winos and the lightest
neutralino $(\lambda)$ reads
\bea\label{lagrangian}
{\cal L}&=&
-{1\over 4} F_{\mu\nu} F^{\mu\nu}
-{1\over 2}W^+_{\mu\nu} W^{-\mu\nu}
+m_W^2W_\mu^+W^{-\mu} \nn \\
&+& {\bar w}(i\gamma^\mu D_\mu -m_w)w+
 {1 \over 2}{\bar \lambda}(i\gamma^\mu\partial_\mu-m_\lambda)\lambda \nn \\
&-&{e \over 2 \sqrt{2} \sin\theta_W}
(W^+_\mu {\bar \lambda}\gamma^\mu (v+a\gamma_5)w+
W^-_\mu {\bar w}\gamma^\mu (v+a\gamma_5)\lambda)
,
\eea
where $F_{\mu\nu}=\partial_\mu A_\nu-\partial_\nu A_\mu$,
$W^-_{\mu\nu}=D_\mu W^-_\nu-D_\nu W^-_\mu$ and $D_\mu=\partial_\mu -ieA_\mu$;
$\theta_W$ is the weak angle, $e$ is the electric charge, $v$ and $a$
are numbers of order
unity which depend on details of the chargino and neutralino
mass matrices. $w$ is a Dirac fermion with the same electric charge
as the $W$-boson and $\lambda$ is a Majorana fermion. For a mass spectrum
satisfying eq. (\ref{lagrangian}) the dominant wino decays are
$w^-\rightarrow W^{-*}\lambda \rightarrow q_1 {\bar q}_2 \lambda$ and
$w^-\rightarrow W^{-*}\lambda \rightarrow l^- {\bar \nu} \lambda$.
For small scalar lepton masses also the decays
$w^-\rightarrow {\tilde l}^{-*}{\bar \nu} \rightarrow l^- {\bar \nu} \lambda$
and
$w^-\rightarrow  l^-{\tilde{\bar \nu}} \rightarrow l^- {\bar \nu} \lambda$
have to be considered.

The total cross section for wino pair production in electron proton
scattering is given by
\beq\label{sigmatot}
\sigma_{tot}^w(s)=\int_0^{0.83}dx_1\int_0^1dx_2f_{\gamma|e}^{LASER}(x_1)
f_{\gamma|p}(x_2)\sigma_{\gamma\gamma}^w(x_1x_2s),
\eeq
where $\sigma_{\gamma\gamma}^w$ is the wino pair production cross section
in $\gamma\gamma$ scattering, $f_{\gamma|e}^{LASER}(x)$ is the spectrum of
the Compton back-scattered laser photons and
$f_{\gamma|p}(x)$ is the spectrum of photons radiated from the proton.

The laser photon spectrum has been calculated by I.F. Ginzburg et al.
\cite{laser}. Contrary to the Weizs\"acker-Williams spectrum it is hard
\cite{laser,zerwas}
\bea
&&f_{\gamma|e}^{LASER}(x)=2{(1+Y)^2 \over (1-x)^2} \times \\
&& \hspace{-1cm}
\frac {2\,Y^2-4\,Y\left (1+Y\right
)x+\left (4+4\,Y+3\,Y^2\right )x^{2}-Y^{2}x^3}
{Y\left (16+32\,Y+18\,Y^2+Y^3\right )-2\,\left (
8+20\,Y+15\,Y^2+2\,Y^3-Y^4\right )\ln (1+Y)}, \nn
\eea
and peaks at the maximum value of $x_{max}=Y/(1+Y)=0.83$
for the optimized value of the parameter $Y=4.82$. The photon spectrum
$f_{\gamma|p}(x)$ has an inelastic component which is obtained
by summing the Weizs\"acker-Williams spectra of the quarks inside the proton
in the case of inelastic proton scattering,
\beq
f_{\gamma|p}^{inel}(x)=\int_0^1dx_1 \int_0^1 dx_2 \sum_q f_{\gamma|q}(x_1)
f_q(x_2,Q^2)\delta(x-x_1x_2);
\eeq
here $f_q(x,Q^2)$ are the parton densities and  $f_{\gamma|q}(x)$
is the Weizs\"acker-Williams spectrum
\beq
f_{\gamma|q}(x)=
e_q^2{\alpha \over 2\pi}{1+(1-x)^2 \over x}\ln{t_{max} \over t_{cut}},
\eeq
where $t_{max}$ and $t_{cut}$ are characteristic maximal and
minimal momentum transfers. Following \cite{altarelli}, we  choose
$t_{max}=\hat s-4m_w^2$ for wino pair production, where $\hat s$
is the photon-quark c.m.s. energy, and $t_{cut}=1\  GeV^2$. Furthermore we
use the Duke-Owens parton densities set 1, with $Q=100\ GeV$
\cite{duke-owens}.

Several years ago, Drees and Zeppenfeld have shown for the case
of scalar electron-photino production in $e^-p$ scattering
that the cross section for the elastic process
$e^-p \rightarrow {\tilde e}\lambda p$ and the inelastic process
$e^-p \rightarrow {\tilde e}\lambda p$ are of the same size \cite{drees}.
This is very interesting, especially since the elastic process
yields a very clean final state. The elastic cross section
can be evaluated using a modified Weizs\"acker-Williams approximation.
An accurate expression for the corresponding photon spectrum
has been derived by Kniehl \cite{kniehl}:
\beq\label{kniehl1}
f^{el}_{\gamma|p}(x)=-{\alpha \over 2\pi}x
\int_{-\infty}^{t_0}{dt \over t}
\left(2\left [{1 \over x}({1 \over x}-1)
+{m_p^2 \over t}\right ]H_1(t)+H_2(t)\right),
\eeq
where $m_p$ is the proton mass and $t_0=-m_p^2x/(1-x)$ . $H_1$ and $H_2$ are
combinations of the electric and magnetic form factors of the proton,
$G_E(t)=(1-t/0.71\ GeV^2)^{-2}$ and $G_M(t)=2.79G_E(t)$,
\bea\label{kniehl2}
&&H_1(t)={G_E^2(t)-(t/4m^2)G_M^2(t) \over 1-t/4m^2}, \nn \\
&&H_2(t)=G_M^2(t).
\eea
{}From eqs. (\ref{kniehl1},\ref{kniehl2}) one can obtain
the explicit expression for
$f^{el}_{\gamma|p}(x)$  which can be found in ref \cite{kniehl}.

In order to obtain the wino pair production cross section (\ref{sigmatot})
we finally need the cross section of the two-photon subprocess \cite{fermion},
\bea\label{sigma_fermion}
&&\sigma_{\gamma\gamma}^w(s)= \\
&&4\,\pi \,\alpha^{2}\left (\left (1+{\frac {4\,m^{2}}{s}}-{\frac {8\,m^
{4}}{s^{2}}}\right )\ln ({\frac {1+\beta}{1-\beta}})-\beta\,\left (1+{
\frac {4\,m^{2}}{s}}\right )\right )s^{-1}, \nn
\eea
where $s>4m_w^2$ and $\beta=(1-4m_w^2/s)^{1/2}$ is the wino velocity.
We shall also need the corresponding cross section for the production of
$W$-boson pairs \cite{Wprod},

\bea\label{sigma_W}
&&\sigma_{\gamma\gamma}^W(s)= \\
&&8\,\pi \,\alpha^{2}\beta\,\left (1+{\frac {3\,m^{2}}{4\,s}}+{\frac {3
\,m^{4}}{s^{2}}}-3\,m^{4}\left (1-{\frac {2\,m^{2}}{s}}\right )\ln ({
\frac {1+\beta}{1-\beta}})s^{-2}\beta^{-1}\right )m^{-2}, \nn
\eea
which, for $m_W<m_w$, is always larger than the
production cross section for  winos.

{}From eqs. (\ref{sigmatot})-(\ref{sigma_fermion}) one now obtains the
elastic and inelastic wino pair production cross sections.
The results are shown in fig. 2 for three different electron-proton c.m.s
energies, $\sqrt{s}=314,\ 450,\ 1000\ GeV$. Note, that the elastic
cross section is slightly larger than the inelastic one. The total cross
sections are plotted on fig. 3. At the HERA energy of $314 \ GeV$ the cross
section is  $10\ fb$ for a wino mass of $60\ GeV$, which is significantly
below the reach of LEP II. At $1\ TeV$ c.m.s. energy the production cross
section is larger than $1\ fb$ for $m_w<190\ GeV$, which corresponds
to the reach of a
linear collider with c.m.s. energy of about $400\ GeV$.

Winos decay predominantly into 2 jets and a neutralino (cf. fig. 1).
Wino pair production therefore leads to final states with
4 jets, missing energy and the proton remnant, a clean signature for
supersymmetry. For the leptonic wino decay,
$w ^-\rightarrow l^-{\bar \nu}\lambda$, the final state would
contain a charged lepton pair $l^+l^-$, missing energy and the
proton remnant. In this case the background from
$W$-boson pair production with leptonic $W$-boson decay is important.
The elastic and inelastic
$W$-boson cross sections can be obtained from eqs.
(\ref{sigmatot})-(\ref{kniehl2}), (\ref{sigma_W}) and are shown in fig.4.

We conclude that high energy photon-proton scattering would allow
to search for super-particles in a very interesting mass range,
which raises the question of the general physics potential of
a photon-proton collider.

We thank D. Zeppenfeld and P. Zerwas for helpful discussions.
Z.F. acknowledges partial support from Hung. Sci. Grant under Contract No.
OTKA-F1041.

\vfill\eject

\vfill\eject

\noindent
{\large {\bf Figure caption}}

\medskip
\noindent
1. Feynman diagram of the wino pair production and dominant decay
in photon-proton collision.

\medskip
\noindent
2. Cross sections of the elastic (solid line) and inelastic (dashed line)
wino pair productions at different c.m.s. energies.

\medskip
\noindent
3. Total cross section of the wino pair productions at different
c.m.s. energies.

\medskip
\noindent
4. The $W^+W^-$ production cross sections at the corresponding energies
(short dashed: elastic contribution, long dashed: inelastic contribution,
solid line: sum).

\vfill\eject
\end{document}